\documentclass[12pt]{iopart}

\usepackage{graphicx}
\usepackage{hyperref}
\usepackage{xspace}
\usepackage{xcolor}
\usepackage{booktabs}
\usepackage{array}
\usepackage{varwidth}
\usepackage{makecell}

\newcommand{\hopaas}{\ensuremath{\mbox{\sc Hopaas}}\xspace}
\newcommand{\version}{\ensuremath{\mbox{\tt version}}\xspace}
\newcommand{\ask}{\ensuremath{\mbox{\tt ask}}\xspace}
\newcommand{\tell}{\ensuremath{\mbox{\tt tell}}\xspace}
\newcommand{\prune}{\ensuremath{\mbox{\tt should\_prune}}\xspace}
\newcommand{\restapis}{\ensuremath{\mathrm{REST~APIs}}\xspace}
\newcommand{\dockercompose}{\ensuremath{\mbox{\tt docker-compose}}\xspace}
\newcommand{\uvicorn}{\ensuremath{\mathrm{Uvicorn}}\xspace}
\newcommand{\fastapi}{\ensuremath{\mathrm{FastAPI}}\xspace}
\newcommand{\optuna}{\ensuremath{\mbox{\sc Optuna}}\xspace}
\newcommand{\nginx}{\ensuremath{\mathrm{NGINX}}\xspace}
\newcommand{\chartist}{\ensuremath{\mbox{\sc Chartist}}\xspace}

\newcommand{\lamarr}{\ensuremath{\mbox{\sc Lamarr}}\xspace}
\newcommand{\marconi}{\ensuremath{\mbox{\sc Marconi 100}}\xspace}

\begin{document}

\title{Hyperparameter Optimization as a Service on INFN Cloud}

\author{Matteo Barbetti$^{1,2}$ and Lucio Anderlini$^2$}

\address{
$^1$ Department of Information Engineering, University of Firenze,\\
\,\,\, via Santa Marta 3, Firenze (FI), Italy\\
$^2$ Istituto Nazionale di Fisica Nucleare, Sezione di Firenze,\\
\,\,\, via G. Sansonse 1, Sesto Fiorentino (FI), Italy
}

\ead{\href{mailto:matteo.barbetti@fi.infn.it}{Matteo.Barbetti@fi.infn.it}}

\begin{abstract}
The simplest and often most effective way of parallelizing the training 
of complex machine learning models is to execute several training 
instances on multiple machines, scanning the hyperparameter space to 
optimize the underlying statistical model and the learning procedure. 
Often, such a meta-learning procedure is limited by the ability of 
accessing securely a common database organizing the knowledge of 
the previous and ongoing trials. Exploiting opportunistic GPUs 
provided in different environments represents a further challenge 
when designing such optimization campaigns. In this contribution, we 
discuss how a set of \restapis can be used to access a dedicated 
service based on INFN Cloud to monitor and coordinate multiple 
training instances, with gradient-less optimization techniques, 
via simple HTTP requests. The service, called \hopaas 
(\emph{Hyperparameter OPtimization As A Service}), is made of a web 
interface and sets of APIs implemented with a \fastapi backend 
running through \uvicorn and \nginx in a virtual instance of 
INFN Cloud. The optimization algorithms are currently based on 
Bayesian techniques as provided by \optuna. A Python frontend is 
also made available for quick prototyping. We present applications 
to hyperparameter optimization campaigns performed by combining 
private, INFN Cloud, and CINECA resources. Such multi-node multi-site 
optimization studies have given a significant boost to the development
of a set of parameterizations for the ultra-fast simulation of the
LHCb experiment.
\end{abstract}

\section{Introduction}
\label{sec:intro}
In the last decade, machine learning~(ML) has become an incredibly 
valuable tool in practically every field of application, from 
scientific research to industry. Increasingly complex models exhibit 
performance unthinkable until few years ago in a wide range of 
applications, such as image generation~\cite{Ramesh:2021}, language
modelling~\cite{Brown:2020} or medical diagnosis~\cite{Richens:2020}.
Most of the ML techniques rely on the optimization of an \emph{objective 
function} with respect to some internal parameters, describing the 
performance of the algorithm. Usually, when the optimum of the objective
function is a minimum, the names \emph{cost} or \emph{loss function} 
are adopted. The fastest iterative optimization techniques rely on 
the (Stochastic) Gradient Descent technique~\cite{Orr:2003}.
Unfortunately, for a wide class of optimization problems the gradient 
of the loss function with respect to the model parameter is extremely
expensive to compute or cannot be defined at all. For example, 
optimization problems involving noisy loss functions in contexts where 
analytical derivatives cannot be computed cannot rely on 
gradient-descent techniques, requiring the adoption of slower, often 
heuristic, methods. A widely adopted option is to define a 
\emph{surrogate model} describing the variations of the loss function 
across the parameter space together with its uncertainty. Such model is
then employed to drive the optimization algorithm to explore those 
regions where improvements were not statistically excluded from 
previous evaluations. Techniques adopting this approach are referred to 
as Bayesian optimization~(BO) methods and have been an active area of 
research in ML for the last decade~\cite{Bergstra:2011, Bergstra:2013, 
Golovin:2017, Liaw:2018, Akiba:2019, Head:2021, Song:2022}.

Tuning the performance of ML models may benefit from 
\emph{hyperparameter} optimization~(HPO). Hyperparameters are defined 
as all those parameters that are not learned during the model training 
procedure, but rather encode some arbitrariness in the architecture of 
the model itself or in the procedure to train it~\cite{Bergstra:2011}.
In practice, HPO studies require training the model multiple times 
to explore the hyperparameter space. Since training ML models is 
computationally expensive, HPO campaigns should focus as much as 
possible on those regions of the hyperparameter space where the model 
performs better. This allows to reduce the time needed for finding the 
best configuration of the model under investigation. Typically, 
multiple training procedures may result in different model performance 
because of the intrinsic randomness of the stochastic gradient-descent
techniques. Namely, the loss is often a noisy function of the 
hyperparameters.

Exploring the hyperparameter space requires many independent trainings, 
or \emph{trials}, that can run in parallel on different computing 
resources. In general, the greater the number of resources tackling 
trainings, the larger the hyperparameter space is explored. This makes
it possible to obtain better models. Opportunistic access to computing 
resources may provide valuable contribution to HPO campaigns. 
Unfortunately, coordinating studies on resources from different 
providers, restrictions, and regulations challenges the adoption 
of existing HPO services.

In this document, we propose \hopaas (\emph{Hyperparameter 
OPtimization As A Service}). \hopaas allows to orchestrate HPO 
studies across multiple computing instances by using a minimal set 
of \restapis. Computing nodes from multiple HPC centers can concur 
\emph{dynamically} to the same optimization study, requesting to the 
\hopaas server a set of hyperparameters to test and then sending back 
the outcome of the training procedure. Several trials of one or more 
studies can be tracked and monitored through the web interface provided 
by the \hopaas service. A reference implementation, with a server 
instance\footnote{Visit \url{https://hopaas.cloud.infn.it} for 
additional details.} deployed on INFN Cloud resources and a simple 
client package~\cite{Barbetti:2023} wrapping the \restapis to Python 
functions, is also discussed.

\section{\hopaas API specification}
\label{sec:api}
We refer to a \emph{trial} as a single training attempt with a 
specific set of hyperparameters to test. A \emph{study} represents 
an optimization session and includes a collection of trials.
In practice, a study is unambiguously defined by the set of 
hyperparameters to optimize, the range of values where searching 
the optimum, and the modality in which this search is carried out 
(e.g., grid search, Bayesian methods~\cite{Bergstra:2011}, or 
evolutionary algorithms~\cite{Tani:2021}).

The core activity of the \hopaas service is to manage distributed 
optimization studies by providing sets of hyperparameters to
requesting computing nodes, the so-called \hopaas clients. The creation,
intermediate updates, and finalization of a trial is controlled from 
the client-side by using a set of \restapis. Such APIs, named \ask, 
\tell, and \prune, implement these actions upon POST HTTP requests
with user authentication based on an \emph{API token} in the request 
path. A minimal description of the \hopaas \restapis is depicted in
Table~\ref{tab:rest-apis} and further detailed in the rest of this
section.

\begin{table}[]
    \centering
    {\small
    \begin{tabular}{c|m{4.8cm}|c|c}
        \toprule
        \thead{\textbf{API}} & \thead{\textbf{Description}} & \thead{\textbf{HTTP method}} & \thead{\textbf{Request path}}\\[1.5mm]
        \hline
        \version & Provides the version of the \hopaas backend. & GET & \texttt{/api/version}\\[1.5mm]
        \hline
        \ask & Creates a new trial, contributing to a new/existing study. The POST body request should include the set of settings to refer unambiguously to a study. The API response contains the hyperparameters to test. & POST & \texttt{/api/ask/{token}}\\[1.5mm]
        \hline
        \tell & Provides the final score of a trial to the backend optimizer chosen for the study. & POST & \texttt{/api/tell/{token}}\\[1.5mm]
        \hline
        \prune & Provides an intermediate score to the backend optimizer. If the study includes a \emph{pruner} strategy, the API response is a boolean value saying whether or not to continue the current trial. & POST & \texttt{/api/should\_prune/{token}}\\[1.5mm]
        \bottomrule
    \end{tabular}
    }
    \caption{Minimal description of the \restapis provided by the \hopaas service.}
    \label{tab:rest-apis}
\end{table}

A computing node ready to test a set of hyperparameters, whether it 
comes from on-premises, Cloud, or HPC resources, will simply need a
network connection with the \hopaas server to take part to an
optimization campaign. In particular, it will query the \hopaas server 
via the \ask API, including in the request body all the information 
needed to define a study unambiguously. The \hopaas server will 
define a new trial, possibly assigning it to an existing study, or 
creating a new one. Once created the trial, the \hopaas server provides 
it with a unique identifier that is included in the HTTP response 
together with the set of hyperparameters to be evaluated for the study. 

Usually, the evaluation of a set of hyperparameters consists of 
training a model defined by those hyperparameters aiming at the 
resulting value of the objective function. The evaluated performance 
metric may correspond to the loss function computed during the training 
procedure but, in general, it can be any numerical score obtained 
processing a given set of hyperparameters. Once the evaluation is 
completed, the computing node will finalize the trial using the \tell 
API, whose body will include the unique identifier of the trial and 
the final evaluation of the objective function. 

The \hopaas server may serve multiple \ask requests from different 
sources, assigning them to one or different studies, while updating 
the surrogate model each time a new evaluation is made available by 
querying the \tell API.
%

Depending on the specific ML algorithm, intermediate evaluations of 
the objective function can be accessed during the training procedure 
and used to abort non-promising trials (\emph{pruning}) without 
wasting computing power to take the training procedure to an end.
Optionally, the computing node may update the \hopaas server with 
intermediate evaluations of the objective function by querying 
the \prune API for monitoring and pruning purposes. The body of a 
\prune request will contain the unique identifier of the trial, the 
intermediate value of the loss function, and an integer number 
encoding the progress of the training procedure, the so-called 
\emph{step}. The HTTP response will indicate whether the study 
should be early terminated, or it is sufficiently likely to result 
in an improvement over the previous tests.

A reference Python frontend was developed aiming at a facilitated access 
to the \hopaas service from Python applications~\cite{Barbetti:2023}.
While Python is a primary choice for many scientific applications, it 
should be noticed that the client simply wraps the \restapis into 
classes and functions, as the \hopaas protocol is designed to be 
language-agnostic, relying on widely adopted web communication standards.
In addition, the \hopaas client is also framework-agnostic since
the evaluation of the objective function for a given set of 
hyperparameters can be implemented with any framework and environment. 

\section{Implementation}
\label{sec:imple}
\begin{figure}[b]
	\centering
	\includegraphics[width=\textwidth, trim= 0 100px 0 100px]{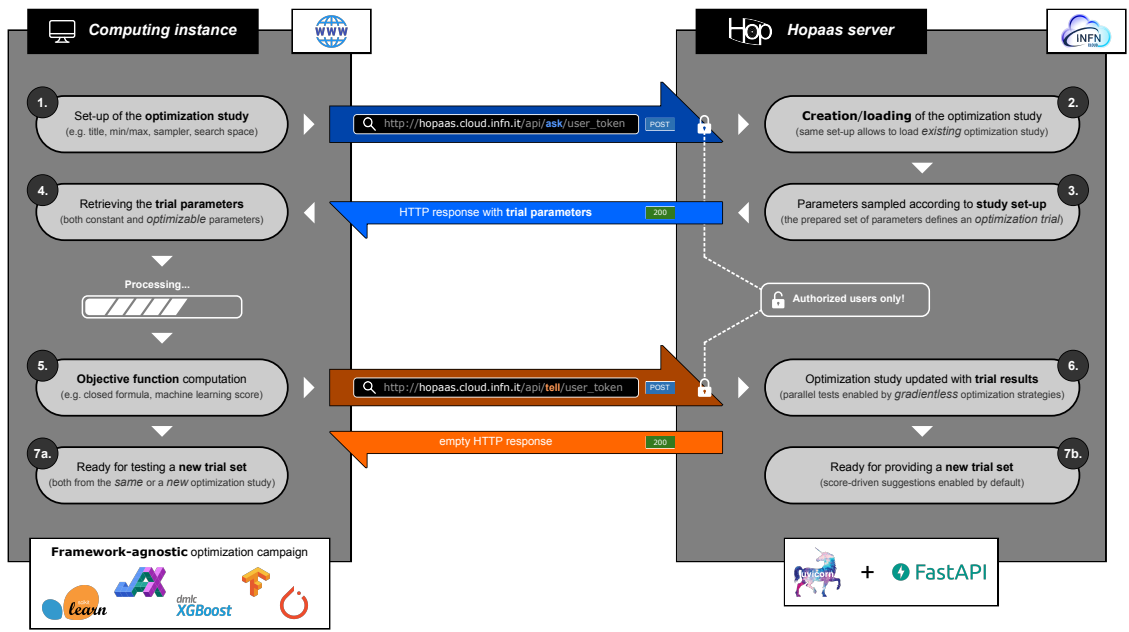}
	\caption{\label{fig:client-server}
		Workflow of an optimization study with a client-server approach based on \restapis. 
	}
\end{figure}

The reference implementation for the \hopaas service running on INFN 
Cloud relies on containerized applications orchestrated with 
\dockercompose~\cite{docker-compose}. The web server implementing 
the \restapis is a scalable set of \uvicorn instances~\cite{uvicorn} 
running an application based on the \fastapi framework~\cite{fastapi}. 
The BO algorithms are provided by integrating the backend 
with \optuna~\cite{optuna}, while future extensions to additional 
frameworks are planned. The access to the \uvicorn instances from the 
Internet is mediated by an \nginx reverse proxy~\cite{nginx} accessed 
via the encrypted HTTPS protocol. A PostgreSQL instance~\cite{postgresql} 
is part of the \dockercompose configuration to provide shared persistency 
to the multiple instances of the web application backend. The workflow 
of the interaction between the \hopaas server and computing nodes is 
depicted in Figure~\ref{fig:client-server}.

The same \hopaas server is designed to serve web-based user access.
A web application, developed in HTML, CSS and JavaScript, is shipped 
to the client browser as defined by a set of web-specific APIs in 
\uvicorn. The web pages of the frontend provide dynamic visualizations 
by fetching data from specialized APIs at regular intervals. Plots 
showing the evolution of the loss reported by different studies 
and trials are obtained with the \chartist library~\cite{chartist-js}.

The user authentication and authorization procedure of the web 
application is managed relying on \emph{access tokens} as defined by 
the OAuth2 standard, using the INFN GitLab instance as identity provider. 
Support for INDIGO IAM is also planned for the future~\cite{Spiga:2020}.
Once authenticated, users can generate multiple API tokens through the 
web application. Each API token has a validity period defined at 
generation and can be revoked at any time. Tokens with shorter validity 
are more appropriate for usage in public or untrusted contexts.

\section{Tuning the LHCb ultra-fast simulation models with \hopaas 
and \marconi}
\label{sec:app}
Machine Learning is an important research area in High Energy Physics 
(HEP), with first applications dating back to the 1990s. Recent years 
have witnessed an explosion in the use of ML techniques in HEP to face 
the computational challenges raised by the upcoming and future 
runs of the Large Hadron Collider (LHC)~\cite{Albertsson:2018}.
With an increasing number, complexity and range of applications of the 
ML models, HPO is becoming popular in HEP~\cite{Wulff:2022}, and 
specialized frameworks targeting distributed computing are being 
developed~\cite{Liaw:2018}.

The reference implementation of the \hopaas service presented here 
has been successfully used for HEP applications and, in particular, 
to optimize the parameterizations for \lamarr~\cite{Anderlini:2022ofl, 
Barbetti:2023bvi}, a novel LHCb \emph{ultra-fast simulation} framework. 
Most of the parameterizations of \lamarr rely on Generative Adversarial 
Networks (GANs)~\cite{Goodfellow:2014}, advanced algorithms taken from 
Computer Vision that were demonstrated to be able to well reproduce the 
distributions obtained from standard simulation 
techniques~\cite{Anderlini:2022ckd, Ratnikov:2023}. Adversarial models 
are particularly sensitive to the choice of the hyperparameter
configuration and require intensive optimization campaigns to model 
accurately the target distributions. 

Several optimization studies have been orchestrated by the \hopaas 
service using \emph{diverse} computing instances, from scientific 
providers (like INFN, CERN and CINECA) and from commercial cloud 
providers (like GCP or AWS). Most of the resources have been
provided by the CINECA supercomputer \marconi, with a custom network 
configuration to enable the communication with the \hopaas 
server~\cite{Mariotti:2021}. \hopaas was able to coordinate dozens 
of optimization studies with hundreds of trials on each 
study from more than twenty concurrent and diverse computing nodes.
This complex setup has allowed to outperform the previous results
and obtain a set of GAN models that succeeds in parameterizing the
high-level response of the LHCb experiment~\cite{Anderlini:2022ofl, 
Barbetti:2023bvi}.

\section{Conclusion and future work}
\label{sec:end}
Hyperparameter tuning and Bayesian methods for gradient-less optimization
provide an effective and simple mean of exploiting opportunistic compute 
resources to improve ML models. Unfortunately, environment variability 
and constraints set by different resource providers make the application 
of existing HPO services challenging. With \hopaas, we propose a solution
designed to require the addition of the thinnest possible layer in the 
model training application, querying a central service via HTTPS and 
minimal \restapis. A reference implementation with a server instance 
running on INFN Cloud and a Python client was presented and tested in 
a real-world application to coordinate hyperparameter optimization 
campaigns on multiple resource providers including INFN, CERN, and 
CINECA. In the future we will improve the quality of the Web User 
Interface, for example enabling custom model documentation and sharing 
among multiple users, and introduce support to multi-objective 
optimizations.

\ack{
We would like to thank Doina Cristina Duma and the rest of the 
INFN Cloud group for the technical support in the deployment 
and test of \hopaas.
We acknowledge enlightening and motivating discussions with Diego 
Ciangottini, Stefano Dal Pra, Piergiulio Lenzi and Daniele Spiga,
especially on future applications and developments.

This work is partially supported by ICSC -- \emph{Centro Nazionale di Ricerca in High Performance Computing, Big Data and Quantum Computing}, funded by European Union -- NextGenerationEU.
}

\section*{References}
\bibliographystyle{bib/iopart-num.bst}
\bibliography{bib/main.bib}

\end{document}